\documentclass[10pt,conference]{IEEEtran}

\usepackage{cite}
\usepackage{amsmath,amssymb,amsfonts}
\usepackage{graphicx}
\usepackage[ruled,vlined,linesnumbered]{algorithm2e}
\DontPrintSemicolon
\SetKwInput{KwInput}{input}
\SetKwInput{KwOutput}{output}
\graphicspath{{figures/}{figures/section3/}{figures/section5/}}
\usepackage{textcomp}
\usepackage{xcolor}
\usepackage{url}
\usepackage{hyperref}
\hypersetup{
  colorlinks=true,
  linkcolor=black,
  citecolor=black,
  urlcolor=black
}
\usepackage{booktabs}
\usepackage{multirow}
\usepackage{enumitem}
\usepackage{microtype}
\usepackage{xspace}
\usepackage{placeins}
\usepackage{tabularx}
\usepackage{array}
\usepackage{framed}
\usepackage{tikz}

\usetikzlibrary{positioning}
\usetikzlibrary{arrows.meta}
\usetikzlibrary{patterns}
\usetikzlibrary{shapes.geometric}

\definecolor{cSus}{HTML}{0072B2}
\definecolor{cUnr}{HTML}{E69F00}
\definecolor{cTie}{HTML}{BBBBBB}
\definecolor{cRev}{HTML}{D55E00}

\newcommand{\framework}{\textsc{ClaimStab-QC}\xspace}
\newcommand{\revcolor}{\color{black}}
\newcommand{\rev}[1]{{\leavevmode\revcolor#1}}

\usepackage[most]{tcolorbox}

\newtcolorbox{rqanswerbox}{
enhanced,
breakable,
colback=black!3,
colframe=black!3,
boxrule=0pt,
borderline west={1.2pt}{0pt}{black!45},
arc=0pt,
left=5pt,
right=4pt,
top=3pt,
bottom=3pt,
before skip=0.6em,
after skip=0.6em
}

\newcommand{\rqanswer}[1]{%
\begin{rqanswerbox}
\textbf{Takeaway.}~#1
\end{rqanswerbox}
}

\begin{document}





\title{When ``A Beats B'' Needs Evidence: \framework{} for Auditing Comparative Claims in Quantum Software}

\author{
\IEEEauthorblockN{
Boshuai Ye\IEEEauthorrefmark{1},
Peng Liang\IEEEauthorrefmark{2},
Maryam Tavassoli Sabzevari\IEEEauthorrefmark{1},
and Arif Ali Khan\IEEEauthorrefmark{1}
}
\IEEEauthorblockA{\IEEEauthorrefmark{1}University of Oulu, Finland\\
Email: \{boshuai.ye, maryam.tavassolisabzevari, arif.khan\}@oulu.fi}
\IEEEauthorblockA{\IEEEauthorrefmark{2}School of Computer Science, Wuhan University, China\\
Email: liangp@whu.edu.cn}
}

\maketitle

\begin{abstract}
Empirical quantum-software papers often report comparative claims: one compiler produces smaller circuits, one optimizer is more reliable, or one ansatz achieves better solution quality. These ``A beats B'' conclusions are not properties of a tool alone; they can change with benchmark scope, circuit construction, compilation, sampling, backend and noise assumptions, optimizer choices, and resource budgets. Existing testing, benchmarking, and reproducibility methods help assess programs, tools, executions, platforms, and experiments, but they do not directly audit whether the reported comparison itself is supported by the evidence exposed in the paper and artifact.

We present \framework{}, a source-bounded framework for auditing empirical comparisons in quantum software. Given a reported comparison, the framework records the compared baselines, metric, relation, and admissible evidence; locks the comparison design before outcomes are computed; and reports either a scoped relation outcome or an explicit evidence boundary. For strict scalar-directional comparisons, the reported direction is classified as Sustained, Unresolved, or \rev{Not preserved} within the locked audit scope.

We evaluate the framework on 455 comparative claims from \rev{a sample of} 119 quantum-software papers. The central finding is a materialization gap: 175 claims can be represented for audit planning, 79 become scalar-directional planning records, and 53 yield lockable audit or diagnostic designs, but only 8 expose enough matched evidence to audit the original comparison without proxy reconstruction. These 8 records yield 2 Sustained, 4 Unresolved, and 2 \rev{Not preserved} outcomes. Controlled diagnostics over 24 benchmark-relevant quantum-workload comparisons further show that simpler checks can preserve apparent directions whose support weakens under locked audit designs. The results suggest that empirical quantum-software sources should report performance orderings together with the matched evidence needed to audit the scope of those orderings.
\end{abstract}

\section{Introduction}
\label{sec:introduction}

Empirical quantum-software papers often report comparative claims: one compiler produces smaller circuits, one optimizer converges faster, one ansatz gives better solution quality, or one backend configuration is more efficient. These claims usually have the form ``A beats B'' under a stated metric. In quantum software engineering (QSE), however, such orderings are not properties of a technique alone. They are produced by a stack of frameworks, circuit libraries, compilers, optimizers, simulators, cloud platforms, and hardware backends~\cite{nielsen2010quantum, shor1994algorithms, grover1996fast, preskill2018quantum, peruzzo2014variational, farhi2014quantum, cirq, bergholm2018pennylane, rohe2025quantum}. The ordering between two baselines can change with benchmark scope, circuit construction, compilation, sampling, backend/noise assumptions, optimizer stochasticity, and resource budget~\cite{li2019tackling, stefano2024empirical}. This makes comparative evidence especially important in QSE: \emph{empirical quantum-software studies need to show not only who wins, but also what evidence supports that ordering and where the support ends.}

Rerunning an experiment and auditing a comparison answer different questions. A rerun checks whether one reported result can be reproduced under one set of conditions; a comparison audit checks whether the stated ``A beats B'' relation is supported by the evidence exposed in the source paper and artifact. Existing methods provide important pieces of this evidence: testing checks programs and platforms~\cite{paltenghi2023morphq}, benchmarking measures tools and backends~\cite{nation2024benchmarking}, reproducibility infrastructure reruns experiments~\cite{mauerer20221}, and hardware-stability studies characterize device variation~\cite{dasgupta2024stability}. However, these studies leave one relation-level object unchecked: the reported comparison itself. \framework{} therefore asks a different question: \emph{whether the stated comparative relation is supported by matched evidence exposed in the source paper or artifact.} Baselines, backend settings, compiler seeds, shot budgets, and outcome rules must come from the source evidence; unsupported choices become evidence boundaries.

\textit{Challenges:} Auditing comparative relations in QSE is difficult for three reasons. (1) The reported winner can depend on the stack: benchmark scope, circuit construction, compilation choices, backend/noise assumptions, and resource policies can change which baseline appears better~\cite{li2019tackling,stefano2024empirical,nation2024benchmarking,dasgupta2024stability}. (2) The audit must stay within the source evidence: baselines, devices, seeds, budgets, and outcome rules must come from the source; unsupported additions create a new comparison. (3) Different claims need different evidence: gate-count, runtime, aggregate-quality, and crossover claims require different paired data and decision rules. These challenges are especially important in QSE because quantum-software comparisons are sensitive to compilation choices, noise, stochastic optimization, and backend variation~\cite{willsch2020benchmarking,baheri2022quantum,dasgupta2023adaptive,dasgupta2024stability}. Thus, a source can expose executable artifacts while still leaving its reported comparison unauditable.

To address these challenges, \framework{} treats the reported comparative relation as the audit unit. It locks source-supported evidence and reports either a scoped relation-level outcome or an evidence boundary. The central constraint is the \emph{materialization gap}: empirical QSE practice may report clear comparative relations without exposing the matched evidence needed for proxy-free auditing.

We apply \framework{} to 455 comparative claims from 119 quantum-software papers. The study exposes a materialization funnel from reported comparisons to auditable evidence: many comparisons can be identified, represented, and planned, but far fewer expose enough matched evidence for proxy-free relation-level auditing. 

This paper makes three \textbf{contributions}:
\begin{itemize}
\item \textbf{Framework.} We present \framework{}, a QSE framework that makes reported empirical comparisons auditable through source-bounded comparison records, locked audit designs, relation-specific outcomes, and explicit evidence boundaries.
\item \textbf{Empirical study.} We apply the framework to 455 claims from 119 papers and reveal a materialization funnel from reported comparisons to auditable evidence, showing where proxy-free auditing is possible and where locked audit designs weaken apparent comparative support.
\item \textbf{Reusable artifact.} We release an audit-evidence package containing claim representations, audit records, locked traces, missing-evidence annotations, and reporting guidance for inspecting empirical quantum-software comparisons~\cite{claimstab_artifact}.
\end{itemize}
\section{Background}
\label{sec:background}

\textit{Quantum computing and the quantum-software stack}: Quantum programs are expressed as circuits of gates acting on qubits, executed on simulators or hardware backends, and read out by measurement~\cite{nielsen2010quantum, preskill2018quantum}. In practice, quantum software is evaluated through a layered stack of programming frameworks, circuit libraries, compilers, optimizers, simulators, cloud platforms, and hardware backends~\cite{cirq, bergholm2018pennylane, rohe2025quantum}. Empirical QSE comparisons therefore evaluate techniques within interacting stack layers, not isolated circuits or algorithms alone.




\textit{Empirical comparisons in quantum software:} Quantum-software papers frequently report that one method performs better, worse, or differently than another on a chosen metric. Examples include compiler or layout claims about two-qubit gates, controlled-NOT (CNOT)/SWAP count, or circuit depth~\cite{li2019tackling,nation2024benchmarking}; aggregate benchmarking claims such as geometric-mean resource reductions across benchmark rows~\cite{kharkov2022arline}; and variational-algorithm claims, including variational quantum eigensolver (VQE) and quantum approximate optimization algorithm (QAOA) claims, about solution quality or resource cost under a chosen ansatz, depth, or configuration~\cite{peruzzo2014variational,farhi2014quantum}. Across these cases, the claim is not only that an experiment ran, but that one baseline outperformed another under stated software-stack conditions. Those conditions define what the claim covers and what requires additional evidence.

\textit{Stack-dependent sources of variation:} An ``A beats B'' result can change when the software stack changes. Prior work shows that qubit mapping, routing, transpilation settings, benchmark suites, backend/noise behavior, and resource policies can affect circuit metrics, tool comparisons, and reported outcomes~\cite{li2019tackling, stefano2024empirical, dasgupta2024stability, kharkov2022arline}. Table~\ref{tab:axis_groups} summarizes the audit-axis groups \rev{that} \framework{} \rev{draws from these studies} to inspect this stack\rev{; each audited comparison instantiates these groups with the settings it reports}. SCOPE selects the evidence points being compared; L1, L2, L3, ALGO, and BUDGET describe how those points are produced or measured. Unsupported or relation-changing conditions are recorded as evidence boundaries.

\begin{table}[t]
\centering
\caption{Audit-axis groups inspected before outcome computation\rev{, defined from prior work~\cite{li2019tackling,stefano2024empirical,dasgupta2024stability,kharkov2022arline}}.}
\label{tab:axis_groups}
\footnotesize
\setlength{\tabcolsep}{3pt}
\renewcommand{\arraystretch}{1.08}
\begin{tabularx}{\columnwidth}{@{}l >{\raggedright\arraybackslash}X >{\raggedright\arraybackslash}p{0.22\columnwidth}@{}}
\toprule
\textbf{Axis} & \textbf{Example inspected dimensions} & \textbf{Audit role} \\
\midrule
SCOPE & problem, circuit, qubit, or graph size; benchmark row; curve or device point; instance distribution & selects compared evidence points \\
\addlinespace
L1 & circuit or ansatz construction, decomposition, routing, layout seed, optimization level, basis gates & construction and compilation choices \\
\addlinespace
L2 & shots per estimate, measurement seed, observable grouping, estimator precision, exact-vs-sampled expectation & sampling and estimation choices \\
\addlinespace
L3 & simulator or hardware backend, noise model, calibration, drift, readout and two-qubit error regimes & backend and noise assumptions \\
\addlinespace
ALGO & optimizer seed, initialization, restart policy, optimizer family & algorithmic stochasticity \\
\addlinespace
BUDGET & iteration count, wall time, function evaluations, total shot budget, stopping rule, timeout, repetitions & resource-policy choices \\
\bottomrule
\end{tabularx}
\end{table}


\textit{Motivating example:} A benchmarking study may report that \rev{Qiskit} produces fewer two-qubit gates than \rev{Tket} on selected circuits~\cite{nation2024benchmarking}. To check that ordering, the reader needs paired rows: for the same circuit, compare the Qiskit and Tket two-qubit-gate counts under the same metric definition, compiler versions, basis gates, optimization level, and seed policy. A Qiskit count from one circuit, optimization level, or seed policy cannot support a Tket comparison reported under another setting. If the paired rows and settings are exposed, the comparison can be audited within that scope; if they are missing, the missing row or setting becomes an evidence boundary.
\section{The \framework{} Framework}
\label{sec:framework}


This section defines the \framework{} audit procedure.

\subsection{Framework Overview}
\label{sec:framework_overview}

Figure~\ref{fig:framework_workflow} summarizes the workflow. A reported comparison is represented as a claim card, mapped to source-supported audit records, locked before outcome computation, and reported as either a relation-level outcome or an evidence boundary. Scalar-directional records receive Sustained, Unresolved, or \rev{Not preserved} outcomes; non-scalar records use the source-supported relation rule when that rule can be locked.

\begin{figure*}[t]
\centering
\begin{tikzpicture}[
font=\footnotesize,
box/.style={draw, rounded corners, align=center, minimum height=14mm,
inner sep=3pt, fill=white},
rep/.style={draw, rounded corners, align=center, minimum height=14mm,
inner sep=3pt, fill=black!6},
arr/.style={-{Stealth[length=2.2mm]}, semithick}
]
\node[box] (s1) {\parbox{23mm}{\centering
\textbf{1.\ Claim card}\par
\scriptsize $\langle P,B_A,B_B,M,R\rangle$}};

\node[box, right=4mm of s1] (s2) {\parbox{23mm}{\centering
\textbf{2.\ Audit record}\par
\scriptsize comparison object}};

\node[box, right=4mm of s2] (s3) {\parbox{23mm}{\centering
\textbf{3.\ Locked design}\par
\scriptsize SCOPE\,$\times$\,ENV\par
\scriptsize before outcomes}};

\node[box, right=4mm of s3] (s4) {\parbox{23mm}{\centering
\textbf{4.\ Records}\par
\scriptsize matched cells\par
\scriptsize rows / curves}};

\node[box, right=4mm of s4] (s5) {\parbox{23mm}{\centering
\textbf{5.\ Outcome}\par
\scriptsize S/U/\rev{NP}\par
\scriptsize or source rule}};

\node[rep, right=4mm of s5] (s6) {\parbox{23mm}{\centering
\textbf{6.\ Report}\par
\scriptsize outcome\par
\scriptsize + evidence boundary}};

\foreach \a/\b in {s1/s2,s2/s3,s3/s4,s4/s5,s5/s6}
\draw[arr] (\a) -- (\b);
\end{tikzpicture}
\caption{The \framework{} workflow. A reported comparison is represented as a claim card, locked before outcome computation, evaluated through comparison records, and reported with an explicit evidence boundary.}
\label{fig:framework_workflow}
\end{figure*}


Four safeguards keep the audit faithful to the source comparison:
\begin{itemize}[leftmargin=*,itemsep=1pt]
\item \textbf{Audit the reported comparison.} Outcomes are scoped to audit records, not source-level correctness. 
\item \textbf{Lock before outcome computation.} Designs, source rows, aggregate rules, device curves, or outcome rules are fixed before labels or statistics are computed.
\item \textbf{Vary only admissible conditions.} Active axes must be source-supported, metric-affecting, and relation-preserving.
\item \textbf{Summarize by the reported relation.} Scalar-directional records are assigned Sustained/Unresolved/\rev{Not preserved} (S/U/\rev{NP}) outcomes using the Wilson decision rule defined in Section~\ref{sec:framework_verdict}; non-scalar records use their locked reported relation rules.
\end{itemize}

\subsection{Claim Cards and Audit Records} 
\label{sec:framework_representation}

Table~\ref{tab:framework_concepts} fixes the audit vocabulary used by \framework{}. Claim cards are needed because a reported ordering is not yet an auditable unit: a single source statement may combine multiple comparator pairs, metrics, scopes, or outcome rules. The card fixes these fields before any verdict is assigned, preventing the audit from conflating the reported comparison with a particular rerun or proxy reconstruction. For example, a compiler claim such as ``Qiskit produces fewer two-qubit gates than Tket'' becomes a claim card; once the source-supported circuits, optimization levels, seeds, baselines, and metric are locked, each circuit--optimization-level--seed setting becomes a matched comparison cell. 

\begin{table}[t]
\centering
\caption{Core \framework{} concepts\rev{, tiers, and the evidence and conclusion for each audited relation type}.}
\label{tab:framework_concepts}
\scriptsize
\begin{tabularx}{\columnwidth}{@{}p{0.30\columnwidth}X@{}}
\toprule
\textbf{Concept} & \textbf{Meaning} \\
\midrule
Reported comparison & A source-paper statement comparing two baselines under a metric, scope, and outcome rule. \\
Claim card & Fixed representation of a reported comparison: problem, baselines, metric, and relation. \\
Audit record & One source-supported comparison unit checked by the framework. \\
Locked audit design & Evidence points and evaluation conditions fixed before outcomes are computed. \\
Matched cell & Paired A/B evaluation under the same locked evidence point and setting. \\
Materialization & Whether the source exposes enough evidence to lock the design or outcome rule without proxy reconstruction. \\
Evidence boundary & Missing or unsupported evidence that limits what can be audited. \\
S/U/\rev{NP} verdict & Scalar-directional outcome: Sustained \rev{(Wilson lower bound $\geq\tau$)}, Unresolved, or \rev{Not preserved (upper bound $\leq 1-\tau$)} within the locked design. \\
\rev{Tier~1 (proxy-free scoped)} & \rev{Scalar-directional audit of the reported comparison on matched cells, with source-supported baselines, metric, settings, and outcome rule; receives S/U/NP.} \\
\rev{Tier~2 (summary)} & \rev{Source-supported output: source-materialized (the source's own reported counts), source-rule summary, extension (instances, metric definition, or baseline differ from the reported comparison), or sensitivity sweep.} \\
\rev{Tier~3 (boundary)} & \rev{The design, evidence object, or outcome rule cannot be locked; no verdict.} \\
\midrule
\multicolumn{2}{@{}l}{\rev{\textit{Relation type: evidence needed $\to$ permitted conclusion}}} \\
\rev{Scalar-directional} & \rev{Matched A/B cells under the same scope and settings (runtime: timing protocol and repetitions) $\to$ S/U/NP over the locked design.} \\
\rev{Aggregate} & \rev{Source rows and the source's aggregation rule with its zero/failure convention $\to$ rule holds or fails on those rows.} \\
\rev{Crossover or regional} & \rev{Per-point curves or devices and the crossover predicate $\to$ concordant or discordant.} \\
\bottomrule
\end{tabularx}
\end{table}

\textit{Claim cards:}
A claim card structures a comparative statement from the source as
\begin{equation}
\mathcal{C} =
\langle \mathcal{P}, \mathcal{B}_A, \mathcal{B}_B, \mathcal{M}, \mathcal{R} \rangle,
\label{eq:five_tuple}
\end{equation}
where $\mathcal{P}$ specifies the problem and evidence universe, $\mathcal{B}_A$ and $\mathcal{B}_B$ identify the compared baselines, $\mathcal{M}$ specifies the metric or measurement object, and $\mathcal{R}$ is the reported comparison. The relation $\mathcal{R}$ may be scalar-directional, aggregate, categorical or equality-based, crossover/regional, runtime-ratio, table-valued, or multi-objective. A claim card yields one audit record when the source comparison is atomic; it yields multiple records when the source comparison requires distinct comparator pairs, metrics, relation rules, evidence scopes, or outcome rules.

For the scalar-directional instantiation used in the evaluation (Section~\ref{sec:eval_rq2}), $\mathcal{R}$ is a scalar-directional ordering $\mathcal{D}\in\{<,>\}$ over $\mathcal{M}$. The problem field $\mathcal{P}$ records the reported comparison context needed to interpret the claim, such as the problem family, instance distribution, benchmark scope, and fixed assumptions stated with the comparison. 



\textit{Audit records:}
An audit record $a$ is the source-supported comparison object on which the relation $R$ is checked. For a scalar-directional record, the locked audit design is the SCOPE\,$\times$\,ENV product
\begin{equation}
\mathcal{A}_a = \mathcal{S}_a \times \mathcal{E}_a ,
\label{eq:record_design}
\end{equation}
predeclared and locked before signed margins or outcomes are measured with $N_a = |\mathcal{A}_a|$ matched cells. A matched comparison record pairs one $\mathcal{B}_A$ and one $\mathcal{B}_B$ evaluation under the same admissible SCOPE point and ENV setting, producing a signed margin and a binary direction-preservation indicator. A design with one comparison record yields only a local preserved, tied, or contradicted outcome; a design with multiple predeclared comparison records supports an aggregate scope-relative outcome.

Not all audit records are scalar-directional. When the source comparison is aggregate, regional, runtime-based, diagnostic,
table-valued, or multi-objective, \framework{} locks the evidence object and outcome rule required by that relation. If the required evidence, statistic, convention, or outcome rule cannot be reconstructed from the available source or artifact evidence, the audit stops with an evidence boundary. \rev{Table~\ref{tab:framework_concepts} maps each audited relation type to the evidence it needs and the conclusion that evidence permits.}

\subsection{Audit Space and Materialization}
\label{sec:framework_audit_space}

\emph{Materialization} asks whether the source exposes enough evidence to lock the comparison object, evidence points, evaluation conditions, and outcome rule. \emph{Auditing} then asks what the locked comparison supports; records that cannot be locked are reported as boundary records.

\textit{Audit space:}
For a given audit record, the admissible audit space is defined as the product of the evidence scope and the evaluation conditions:
\begin{equation}
\mathcal{S}_{\textit{audit}} =
\mathcal{S}_{\textit{scope}} \times \mathcal{E}_{\textit{env}}.
\label{eq}
\end{equation}
$\mathcal{S}_{\textit{scope}}$ contains the evidence points being compared, while $\mathcal{E}_{\textit{env}}$ contains the conditions under which those points are produced or evaluated. Table~\ref{tab:axis_groups} defines the six audit-axis groups used to inspect this space: SCOPE selects the evidence points, and ENV groups L1, L2, L3, ALGO, and BUDGET describe construction, execution, estimation, optimization, and resource-policy choices.


\textit{Inspection and activation:}
For each audit record involving circuit construction, simulation, or execution, \framework{} inspects the six audit-axis groups in Table~\ref{tab:axis_groups} before measuring outcomes. Inspection is broader than variation: unmentioned or unsupported axes are recorded but not varied. An axis is activated as a varying dimension only when the proposed variation is source-supported, metric-affecting, and relation-preserving. All other axes are kept fixed, confined to the source-exposed setting, marked inadmissible, or recorded as insufficiently supported/evidence-boundary cases, together with the source pointer or missing-evidence note that justifies the decision.

\textit{Materialization test:}
For scalar-directional audit records, materialization requires pairable baseline records under the same cell setting. For relation-specific audit records, such as aggregate, crossover, runtime, or table-valued records, it requires the source rows, aggregate rule, curve definition, timing convention, table rule, or regional outcome rule. If these cannot be locked from the available source or artifact evidence, the auditor does not infer missing values, introduce proxy baselines, or replace the source comparison with a new benchmark choice. \framework{} instead returns a boundary record that names the missing evidence, records the inspected sources, and carries no audit verdict. Records that pass this test proceed to the admissibility rules below; RQ1 applies the test at the corpus scale (Section~\ref{sec:eval_rq1}).

\textit{Admissibility rules:}
To keep variations within the audited comparison, admissible variations must satisfy three conditions:
\begin{enumerate}[leftmargin=*,itemsep=1pt]
\item \textbf{Source support}: the variation is stated by the source, exposed in the artifact, implied by the reported experimental design, or corresponds to a standard execution degree of freedom needed to reproduce the comparison.
\item \textbf{Metric relevance}: the variation can affect the audited metric $\mathcal{M}$ or the measured evidence object.
\item \textbf{Relation preservation}: the variation does not change the compared baselines, metric, relation type, or intended interpretation of the source comparison.
\end{enumerate}

In practice, admissible variations are supported in three ways: the problem or benchmark scope implied by the claim, the evaluation conditions that the claim depends on, and standard-practice degrees of freedom, such as transpiler seed or optimization level, which are metric-affecting and relation-preserving. Excluded axes are still recorded with reasons; for example, a compiler/transpiler audit record may admit transpiler seed, optimization level, and basis-gate variation within the stated family, but excludes adding comparator tools or switching to a runtime metric for a gate-count claim. Runtime and timing repetitions are treated as measurement-protocol ENV settings.

\subsection{Dry-Run-First Audit Procedure}
\label{sec:framework_procedure}

\framework{} enforces lock-before-observing through the dry-run-first procedure in Algorithm~\ref{alg:audit}. The procedure first checks whether the audit record can be materialized, labels admissible and inactive axes, forms and filters a candidate design, and emits a dry-run report containing schemas, counts, exclusions, and missing evidence. Only after this dry run is the design locked and the outcome computed. 

\begin{algorithm}[t]
\small
\KwInput{audit record $a$ from claim card $\mathcal{C}$; source/artifact evidence; threshold $\tau$; audit budget $N_{\max}$}
\KwOutput{scoped outcome $O$, explanation report $X$, and evidence boundary $B$}

$r \leftarrow \textsc{RelationType}(a)$;
$(ok,B) \leftarrow \textsc{CheckMaterialization}(a,\textit{source},\textit{artifact})$;

\If{\textbf{not} $ok$}{
\Return{$\textsc{BoundaryOnly}(B)$};
}

$(axes,B) \leftarrow \textsc{LabelAxes}(a,\textit{source},\textit{artifact})$;
$G \leftarrow \textsc{FormCandidateDesign}(a,axes)$;
$(V,B) \leftarrow \textsc{ApplyValidityFilters}(G,B)$;
$\textsc{EmitDryRun}(V,axes,B)$;

\If{$V=\emptyset$}{
\Return{$\textsc{BoundaryOnly}(B)$};
}

$A \leftarrow \textsc{LockDesign}(V,N_{\max})$;

\eIf{$r=\textsc{ScalarDirectional}$}{
$\Delta \leftarrow \textsc{MeasureMargins}(A,\mathcal{B}_A,\mathcal{B}_B,\mathcal{M})$;
$(k,N) \leftarrow \textsc{PreservationCount}(\Delta,\mathcal{R})$;
$O \leftarrow \textsc{Wilson\rev{Rule}}(k,N,\tau)$;
}{
$O \leftarrow \textsc{RelationSpecificOutcome}(A,r,\mathcal{R})$;
}

$(X,B) \leftarrow \textsc{ExplainOutcome}(O,A,axes,B)$;
\Return{$(O,X,B)$};

\caption{\framework{} dry-run-first audit}
\label{alg:audit}
\end{algorithm}

\textit{Audit-budget handling:}
The comparison-record count $N$ is the size of the locked design: the full SCOPE\,$\times$\,ENV product supported by source when it fits the audit budget $N_{\max}$, or an outcome-blind predeclared subset over the same axes when the product exceeds $N_{\max}$. The subset rule, random seed, valid counts, exclusions, and missing fields are reported in the dry-run summary before outcome computation. For a non-scalar record, the outcome rule is applied once its statistic, convention, and rule are locked; otherwise the record returns a boundary record.

\subsection{Scalar-Directional Outcome Assignment}
\label{sec:framework_verdict}

Scalar-directional records need a relation-level rule for deciding whether the reported direction is preserved across the locked design. \framework{} assigns one of three scope-relative outcomes: Sustained, Unresolved, or \rev{Not preserved}. Other relation types use their locked source-supported rules.

\textit{Signed margin and preservation:}
For scalar-directional records, \framework{} treats direction preservation as the event being audited. In each locked comparison record, \framework{} records the signed margin
\begin{equation}
\delta_c = \mathcal{M}_A^{(c)} - \mathcal{M}_B^{(c)} ,
\label{eq:signed_margin}
\end{equation}
whose sign determines whether the claimed direction $\mathcal{D}$ is preserved. For $N$ comparison records with $k$ preserving the direction, the empirical preservation rate is $\hat{s}=k/N$. Because the audited relation is strict, ties do not preserve the relation; they are reported separately from strict contradictions to distinguish tie-driven loss of improvement from opposite-direction dominance. This tie policy applies to strict directional relations such as ``lower'', ``fewer'', or ``better''. Non-strict relations, such as equivalence or no-worse-than claims, are handled by their reported relation rules.

\textit{Wilson S/U/\rev{NP} rule:}
The Wilson rule~\cite{wilson1927probable,brown2001interval} is a transparent audit policy over locked comparison records: a relation is Sustained when the Wilson lower bound reaches $\tau$, \rev{Not preserved} when the Wilson upper bound \rev{is at most} $1-\tau$, and Unresolved otherwise. The interpretation unit is the locked design: the preservation count $k/N$ and S/U/\rev{NP} label are read relative to the locked metric, preservation threshold, confidence level, and comparison records. The preservation threshold and confidence level are predeclared separately. We set $\tau=0.95$ as a conservative threshold for strict directional claims. We report scalar-directional results using Wilson intervals at the commonly used 95\% confidence level~\cite{altman2000statistics}, with bounds $\underline{w}$ and $\bar{w}$. Because locked cells may share circuits, instances, seeds, optimization levels, timing groups, or environments, RQ4 (Section~\ref{sec:eval_rq4}) checks sensitivity to alternative $\tau$ values and dependence-aware choices. \rev{The rule is a decision policy over a finite locked design, and its labels describe that design. At $N=80$, the size of all eight Tier-1 designs, it reduces to counts: Sustained iff $k=80$, Not preserved iff $k=0$, and Unresolved otherwise.} Formally, \framework{} assigns:

\begin{itemize}[leftmargin=*]
\item \textbf{Sustained}: $\underline{w}\geq\tau$.
\item \textbf{\rev{Not preserved} within audited scope}: $\bar{w}\leq 1-\tau$\rev{; ties and contradictions are reported separately, because ties alone can reach this bound}.
\item \textbf{Unresolved}: neither boundary holds.
\end{itemize}

Each verdict is therefore interpreted within the locked audit scope.

\subsection{Explanation and Diagnostic Reports}
\label{sec:framework_explanation}

After assigning the primary scalar-directional outcome, \framework{} records robustness checks, explanation fields, and optional diagnostics. These reports are scope-relative, with explicit evidence boundaries, and leave the primary outcome unchanged.

\textit{Robustness and sensitivity checks:}
The primary scalar-directional outcome uses the locked design and the Wilson S/U/\rev{NP} rule defined in Section~\ref{sec:framework_verdict}. \framework{} then reports secondary checks when applicable: dependence-aware grouping checks, threshold sensitivity over predeclared $\tau$ values, tie-handling sensitivity for strict scalar-directional records with per-cell traces, budget or seed-count expansion while keeping the locked comparison design fixed, and practical-effect and ALGO-bookkeeping checks when the required evidence is available. Missing traces or unsupported checks are recorded as evidence boundaries. These checks qualify the robustness of the primary outcome without changing the predeclared primary verdict.

\textit{Explanation report:}
The scalar explanation report separates \emph{statistical and policy uncertainty} from \emph{audit-axis and evidence-boundary information}. Statistical and policy uncertainty includes the Wilson interval, cluster-resampling check, sensitivity to threshold, audit budget, and practical-effect threshold. Audit-axis and evidence-boundary information records preservation, ties, strict contradictions, active axis groups, inactive or source-confined axes, inadmissible axes, and evidence boundaries. Because the audited direction is strict, ties are non-preserving; the report therefore names the \emph{explanation mode} behind a verdict, and in particular records a \rev{Not preserved} verdict as \emph{strict-contradiction-dominated} when opposite-direction margins dominate or as \emph{tie-dominated} when strict improvement is lost mainly through ties.

\textit{Diagnostic extensions:}
Two optional diagnostics refine or stress-test scalar-directional verdicts without changing the predeclared primary audit. First, when a defensible practical-effect threshold $\epsilon$ is available, \framework{} can replace the basic preservation indicator with
\begin{equation}
I_c^{\epsilon} =
\mathbb{1}\!\left[
\delta_c \text{ preserves } \mathcal{R}
\;\land\;
|\delta_c| \ge \epsilon
\right],
\end{equation}
so a comparison record supports the claim only when its signed margin both preserves the reported relation and exceeds the practical threshold. Wilson classification can then be recomputed using $k_{\epsilon}=\sum_c I_c^{\epsilon}$ and $N=|\mathcal{A}|$.

Second, for an active audit axis $x$, \framework{} reports marginal preservation summaries
\begin{equation}
s_x(v)=
\frac{1}{|\mathcal{A}_{x=v}|}
\sum_{c\in \mathcal{A}_{x=v}} I_c ,
\end{equation}
where $\mathcal{A}_{x=v}$ is the subset of locked comparison records whose axis value is $v$, and $I_c$ is the primary preservation indicator. These summaries identify axis values associated with changes in preservation or signed margins. They are descriptive diagnostics reported alongside the primary locked-design verdict.

\section{Implementation and Artifact}
\label{sec:implementation}

\framework{} is implemented in Python using Qiskit~\cite{javadi2024quantum}, NumPy~\cite{harris2020array}, and SciPy~\cite{virtanen2020scipy}. The artifact records the toolchain used to produce the released audit evidence and includes lightweight checks for consistency of the committed evidence package~\cite{claimstab_artifact}. The implementation separates shared framework logic from record-specific adapters through three layers: (1) a claim registry and materialization checker store claim-card specifications and determine whether the required source evidence can be locked; (2) a locking layer freezes comparison records, axis labels, seeds, baseline measurements, exclusions, and evidence boundaries before outcome computation; and (3) relation checkers compute scalar-directional S/U/\rev{NP} verdicts or relation-specific summaries, while an exporter writes claim cards, locked evidence manifests, per-cell or aggregate traces, outcome rules, explanation fields, and boundary records.

Record-specific adapters translate source formats, source rows, recovered comparisons, or executable comparators into locked audit designs when the required evidence is available. They do not redefine baselines, metrics, admissible axes, or outcome rules; they connect source-supported evidence to the shared audit logic.

\section{Evaluation}
\label{sec:evaluation}

We evaluate \framework{} as a source-bounded procedure for materializing and auditing empirical comparative relations. The evaluation is organized around four research questions (RQs): (1) whether published comparative claims can be materialized for relation-level auditing; (2) what auditing materialized comparison records reveal; (3) whether \framework{} explains where comparative support is lost; and (4) whether verdicts are robust to statistical and declared audit choices. The artifact provides the locked evidence package used for inspection, including claim cards, audit records, released per-cell or aggregate traces, manifest files, verification scripts, and missing-evidence annotations~\cite{claimstab_artifact}.



\subsection{RQ1: Can published comparative claims be materialized for relation-level auditing?}
\label{sec:eval_rq1}

\textit{Corpus construction and screening:}
We constructed the corpus through a structured Google Scholar search-and-screen protocol. We used Google Scholar because it provides broad scholarly discovery across publication venues, preprint repositories, and other research sources. The search was run on January 28, 2026 using 13 predefined query strings that combined quantum-software terms (\texttt{quantum compiler}, \texttt{quantum compilation}, \texttt{transpiler}, \texttt{QAOA}, \texttt{VQE}, \texttt{variational quantum}, and \texttt{quantum software}) with comparison terms (\texttt{benchmark}, \texttt{comparison}, and \texttt{ansatz comparison}). Results were restricted to 2020--2024 using Google Scholar year filters and source metadata during screening, recorded as per-query result \rev{counts}, deduplicated by title and available identifiers such as DOI or arXiv ID, and capped at 250 candidate papers to form a fixed five-year sampled corpus.

We included papers \rev{on compilation/transpilation or variational optimization} that reported at least one quantitative comparison \rev{between two approaches}. We excluded papers that were outside \rev{these two paradigms}; hardware-only, pulse-level-only, error-correction-only, algorithmic-complexity-only, survey-only, or application-only without a software-stack comparison; \rev{or} did not report a quantitative comparative claim. \rev{Of the 250 candidates, 131 were excluded: other paradigms (66), application-only (31), hardware-only (15), no quantitative comparison (9), complexity-only (3), pulse-level (3), error correction (2), and surveys (2).} Screening retained 119 papers; after claim extraction and claim-level correction, RQ1 uses 455 accepted comparative claims. The artifact records the \rev{normalized} query strings, search date, per-query result counts, screening decisions, exclusion reasons, \rev{the 457} full-text extraction records, \rev{the 455 accepted claims, the Tier-1, Tier-2, and boundary audit records,} and \rev{the RQ3 diagnostic grids}~\cite{claimstab_artifact}. \rev{Search results and the candidate and deduplication steps are archived as counts.}

\textit{Claim extraction and coding validity:}
From the retained corpus, we extracted comparative claims through a three-stage worksheet protocol (candidate $\to$ verbatim-verified $\to$ accepted), summarized with the materialization funnel in Figure~\ref{fig:funnel_barriers}. \rev{A claim unit is one comparative statement by the source authors with fixed baselines, metric, direction, and problem scope; restatements were merged, and different comparator pairs, metrics, or problem families were kept separate. A candidate claim was accepted if the source authors themselves report it as a comparison of two approaches on a reported metric or outcome; secondary comparisons that met this test were kept. Of 630 candidates, 628 matched the source text verbatim; deduplication and acceptance gave 457 extraction records, and two source-grounded corrections gave 455 accepted claims.} Each accepted claim is then checked at three increasingly strong layers. Representation checks whether the source text fills the claim-card fields $\langle\mathcal{P},\mathcal{B}_A,\mathcal{B}_B,\mathcal{M},\mathcal{R}\rangle$ (Section~\ref{sec:framework_representation}). Planning checks whether the comparator pair, metric, relation type, and admissible evidence scope can be specified from the available support. Materialization checks whether the required evidence can be locked without proxy reconstruction (Section~\ref{sec:framework_audit_space}), including matched-cell requirements for scalar-directional records and aggregation rule, source universe, comparator evidence, and zero/failure convention for non-scalar records. A runnable artifact can still fail materialization if it exposes code but not these fields.

Coding followed a shared worksheet protocol: one annotator performed the initial annotation, a second annotator independently recoded the items, and a third annotator conducted a stratified 120-item consistency audit over the main auditability-funnel boundaries. Disagreements were adjudicated against the source paper, coding worksheet, and auditability definitions. Agreement was high for the consequential decisions: 92.11\% for planning feasibility ($\kappa = 0.835$, $n = 114$), 96.55\% for auditable-design classification ($\kappa = 0.782$, $n = 58$), and 100.0\% for proxy-free scoped vs. non-exact classification ($\kappa = 1.000$, $n = 58$). \rev{Once the required evidence is identified, this last decision is largely mechanical, so its $\kappa$ of 1 measures how consistently the gate is applied. The earlier decisions need more judgment, as their lower $\kappa$ values show.} The proxy-free scoped count and all eight Tier-1 records remained unchanged; supporting agreement records are provided in the artifact~\cite{claimstab_artifact}.


\textit{Auditability surface:}
Applying this protocol to the accepted claims yields the funnel in Figure~\ref{fig:funnel_barriers}. Of the 455 accepted comparative claims, 175 can be represented as claim cards. Among these, 145 state strict scalar-directional relations and 93 are planning-feasible; their intersection yields 79 scalar-directional planning records. Of these 79 records, 53 yield a lockable audit or diagnostic design, and 8 of them are proxy-free scoped audit records of the reported comparison itself, without substituting unsupported evidence, settings, or baselines.

\begin{figure}[t]
    \centering
    \begin{tikzpicture}[
        font=\scriptsize,
        band/.style={draw, rounded corners, align=center, minimum height=5.4mm,
            text width=66mm, fill=white},
        leaf/.style={draw, rounded corners, align=center, minimum height=5.4mm,
            text width=32mm, fill=white},
        arr/.style={-{Stealth[length=1.8mm]}, black!55, semithick},
        node distance=2.6mm
    ]
        \node[band, fill=black!4] (extract)
            {Extraction: 119 papers $\to$ 630 candidates $\to$ 628 verbatim-verified (99.7\%) \rev{$\to$ 457 records}};
        \node[band, below=of extract, fill=black!8] (accept)
            {455 accepted comparative claims (142 compilation / 313 variational)};
        \node[band, below=of accept, text width=58mm] (spec)
            {175 represented as claim cards};
        \node[band, below=of spec, text width=62mm] (rec)
            {79 scalar-directional planning records (145 scalar $\cap$ 93 planning-feasible)};
        \node[leaf, below=6mm of rec, xshift=-22mm] (exact)
            {8 proxy-free scoped audit records};
        \node[leaf, below=6mm of rec, xshift=22mm, dashed] (design)
            {53 lockable designs (8 scoped + 2 source-materialized + 43 extension)};
        \draw[arr] (extract) -- (accept);
        \draw[arr] (accept) -- (spec);
        \draw[arr] (spec) -- (rec);
        \draw[arr] (rec) -- (exact);
        \draw[arr, dashed] (rec) -- (design);
    \end{tikzpicture}
    \caption{Corpus extraction and materialization funnel. The 53 lockable designs are registry-level design instances; the 8 proxy-free scoped audit records without proxy reconstruction form the Tier-1 RQ2 subset.}
    \label{fig:funnel_barriers}
\end{figure}


The 79-to-8 drop is the materialization-gap result. Table~\ref{tab:materialization_disposition} categorizes the 79 scalar-directional planning records by the strongest evidence that could be locked. The 53 lockable cases consist of 8 proxy-free scoped audits of the reported comparison, 43 source-supported extension or diagnostic designs, and 2 source-materialized cases without matched A/B source data. The remaining 26 records stop at a boundary because the available evidence supports only a source-stated rule or aggregate summary, a verification-only check, or an incomplete timing/comparator convention. The PauliOpt runtime claim illustrates the timing-boundary path; the OLSQ-versus-tket source-table evidence illustrates a source-materialized audit with bounded rerun-configuration evidence~\cite{meijer2023comparison,tan2020optimal}. \rev{Making transpiler seed and optimization level admissible for gate-count and depth claims leaves the count at 8: Section~\ref{sec:framework_audit_space} already admits seed and optimization-level variation when it is source-supported, metric-affecting, and relation-preserving. The other 71 planning records are 43 extension or diagnostic designs, 2 source-materialized records, and 26 boundary records; each has a recorded blocker other than missing hyperparameter, optimizer, or budget settings. Counting the 2 source-materialized records as proxy-free would give 10.}

\begin{table}[t]
\centering
\caption{Materialization breakdown of the 79 scalar-directional planning records. Counts are record-level and sum to 79; recorded reasons summarize the strongest lockable evidence object or the blocking evidence boundary.}
\label{tab:materialization_disposition}
\small
\setlength{\tabcolsep}{3.5pt}
\renewcommand{\arraystretch}{1.05}
\begin{tabular}{p{0.31\linewidth}p{0.50\linewidth}r}
\toprule
\textbf{Category} & \textbf{Recorded reason} & \textbf{Records} \\
\midrule
Proxy-free scoped & Lockable matched comparison evidence; eligible for Tier-1 S/U/\rev{NP} verdicts & 8 \\
Lockable, not proxy-free scoped & Source-supported extension or diagnostic design; not the reported comparison itself & 43 \\
Lockable, not proxy-free scoped & Source-materialized evidence without matched A/B source data for proxy-free scoped auditing & 2 \\
Boundary & Source-rule-only evidence; no matched scalar A/B design lockable for proxy-free scoped auditing & 20 \\
Boundary & Verification-only record, not an auditable design & 2 \\
Boundary & Missing timing protocol, comparator implementation, or other lockable source convention & 4 \\

\bottomrule
\end{tabular}
\end{table}

\rqanswer
{Proxy-free comparative auditing requires evidence that many papers leave implicit; empirical QSE papers should report the baselines, settings, instances, budgets, and outcome rules needed to lock the comparison.}

\subsection{RQ2: What does auditing materialized comparison records reveal?}
\label{sec:eval_rq2}

RQ2 reports audit outputs by evidence tier. \rev{Table~\ref{tab:framework_concepts} defines the three tiers}. Table~\ref{tab:external_audit} reports these records, and the 53-design auditability inventory is provided in the artifact~\cite{claimstab_artifact}.

\textit{Tier-1 eligibility gate:}
The Tier-1 set contains records for which scalar-directional auditing is possible without changing the reported comparison or introducing unsupported baselines, metrics, settings, or outcome rules. A record enters Tier 1 when it (1) states a scalar-directional relation, (2) can be materialized without proxy reconstruction, (3) pairs the two reported comparators under matched comparison cells or a source-supported aggregate object, and (4) admits a locked design before outcomes are observed. Eight audit records meet these requirements and form the proxy-free scoped subset used for S/U/\rev{NP} verdicts. 

\textit{Tier-1 proxy-free scalar-directional audit records:}
Each Tier-1 record uses a primary locked design over source-supported axes. For records with per-cell evidence, $N$ denotes the number of locked comparison records in that design, not a post-hoc sampling target. The primary verdicts cover the full source-supported primary design in each of the eight Tier-1 audits. Seed and repetition axes are finite locked budgets, not enumerations of all possible executions.

Figure~\ref{fig:external_forest} shows the primary locked Wilson 95\% intervals for the eight Tier-1 records. EX-C1 and EX-C2 are Sustained, EX-C3--EX-C6 are Unresolved, and EX-C7 and EX-C8 are \rev{Not preserved} within their audited scopes. Both records preserve the reported direction in $0/80$ primary locked cells\rev{: EX-C8 has 80 contradicted cells, and EX-C7 has 62 contradicted and 18 tied}. The artifact provides source links, audit manifests, locked traces, replay traces, worksheet identifiers, and aggregate/cluster traceability boundaries for EX-C2 and EX-C4~\cite{claimstab_artifact}.

\begin{figure}[t]
    \centering
    \begin{tikzpicture}[x=5.4cm, font=\scriptsize,
      ival/.style={black!70, line width=0.8pt},
      mcirc/.style={circle, draw=black, fill=black!55, inner sep=1.4pt},
      msq/.style={rectangle, draw=black, fill=black!55, inner sep=1.4pt},
      mtri/.style={regular polygon, regular polygon sides=3, draw=black, fill=black!55, inner sep=1.3pt}]
      \draw[black!30, dashed, line width=0.5pt] (0.05,0.25) -- (0.05,-2.95);
      \draw[black!30, dashed, line width=0.5pt] (0.95,0.25) -- (0.95,-2.95);
      \node[black!55] at (0.05,0.5) {$1{-}\tau$};
      \node[black!55] at (0.95,0.5) {$\tau$};
      \draw[black!50, line width=0.5pt] (-0.02,-2.95) -- (1.02,-2.95);
      \foreach \t/\lab in {0/0, 0.25/0.25, 0.5/0.50, 0.75/0.75, 1/1.00}
        \draw[black!50] (\t,-2.95) -- ++(0,-0.05) node[below, black] {\lab};
      \node[font=\footnotesize] at (0.5,-3.55) {Preservation rate, $k/N$};
      \foreach \y/\lo/\hi/\sh/\shp/\lab in {%
        0/0.954/1/1/mcirc/EX-C1,
        -0.38/0.954/1/1/mcirc/EX-C2,
        -0.76/0.672/0.853/0.775/msq/EX-C3,
        -1.14/0.453/0.666/0.5625/msq/EX-C4,
        -1.52/0.441/0.654/0.55/msq/EX-C5,
        -1.90/0.323/0.534/0.425/msq/EX-C6,
        -2.28/0/0.046/0/mtri/EX-C7,
        -2.66/0/0.046/0/mtri/EX-C8}{
          \draw[ival] (\lo,\y) -- (\hi,\y);
          \draw[ival] (\lo,\y) ++(0,0.06) -- ++(0,-0.12);
          \draw[ival] (\hi,\y) ++(0,0.06) -- ++(0,-0.12);
          \node[\shp] at (\sh,\y) {};
          \node[left, black] at (-0.03,\y) {\lab};
      }
      \node[mcirc] at (0.07,-4.0) {};
      \node[right] at (0.085,-4.0) {Sustained};
      \node[msq] at (0.40,-4.0) {};
      \node[right] at (0.415,-4.0) {Unresolved};
      \node[mtri] at (0.73,-4.0) {};
      \node[right] at (0.745,-4.0) {\rev{Not preserved}};
    \end{tikzpicture}
    \caption{Wilson 95\% confidence intervals for the eight Tier-1 proxy-free scoped audit records\rev{; marker shape gives the} primary locked verdict ($\hat{s}=k/N$; primary locked values in Table~\ref{tab:external_audit}).}
    \label{fig:external_forest}
\end{figure}

\textit{\rev{Worked example}, EX-C7:}
EX-C7~\cite{nation2024benchmarking} illustrates relation-level auditing beyond a measurement rerun for a compiler comparison\rev{; Figure~\ref{fig:claim_card} follows this record from source text to outcome}. \framework{} reconstructs the reported Qiskit-versus-Tket ordering as a claim card, locks the connectivity-stress scope, and evaluates four \rev{circuits on a four-node line} across four Qiskit optimization levels and five transpiler seeds \rev{against Tket's default pipeline}. The resulting 80 matched cells contain 0 preserved, 62 contradicted, and 18 tied comparisons (Figure~\ref{fig:exc7_heatmap}). \rev{All 20 cells at level 0 are contradicted; levels 1--3 each have 14 contradicted and 6 tied cells.} Because the reported relation is strict, ties are non-preserving in the primary rule; RQ4 reports tie-handling sensitivity separately. The outcome is a \rev{Not preserved} verdict within the audited scope, driven by strict contradictions, and the locked seed-count expansion yields $0/320$ preserved cells.

\begin{figure}[t]
\centering
\revcolor
\footnotesize
\begin{tabular}{@{}>{\raggedright\arraybackslash}p{0.24\columnwidth} >{\raggedright\arraybackslash}p{0.68\columnwidth}@{}}
\toprule
\textbf{Stage} & \textbf{Value} \\
\midrule
Source text & ``as a whole, Qiskit outperforms BQSKit, Staq, and Tket in terms of 2Q gate count'' \\
Claim card & $\langle P, B_A, B_B, M, R\rangle$ \\
\quad $P$ & transpilation resource benchmarking \\
\quad $B_A, B_B$ & Qiskit; Tket (BQSKit, Staq dropped) \\
\quad $M$; $R$ & two-qubit gate count; $B_A < B_B$ \\
Audited scope & four connectivity-stress circuits, four-node line \\
Locked design & 4 circuits $\times$ 4 Qiskit opt.\ levels $\times$ 5 seeds $=80$ cells; Tket default pipeline \\
Matched cells & 0 preserved, 62 contradicted, 18 tied \\
Outcome & \textbf{Not preserved} within audited scope; $0/320$ under seed expansion \\
\bottomrule
\end{tabular}
\caption{EX-C7 from source text to outcome~\cite{nation2024benchmarking}.}
\label{fig:claim_card}
\end{figure}

\begin{figure}[t]
    \centering
    \begin{tikzpicture}[x=11mm,y=7mm,font=\scriptsize]
      \foreach \r/\c/\co/\ti/\col/\tc in {
        0/0/5/0/{black!60}/{white}, 0/1/4/1/{black!50}/{white}, 0/2/4/1/{black!50}/{white}, 0/3/4/1/{black!50}/{white},
        1/0/5/0/{black!60}/{white}, 1/1/0/5/{black!12}/{black}, 1/2/0/5/{black!12}/{black}, 1/3/0/5/{black!12}/{black},
        2/0/5/0/{black!60}/{white}, 2/1/5/0/{black!60}/{white}, 2/2/5/0/{black!60}/{white}, 2/3/5/0/{black!60}/{white},
        3/0/5/0/{black!60}/{white}, 3/1/5/0/{black!60}/{white}, 3/2/5/0/{black!60}/{white}, 3/3/5/0/{black!60}/{white}}{
          \fill[\col] (\c,-\r) rectangle ++(0.96,0.96);
          \draw[black!30,line width=0.3pt] (\c,-\r) rectangle ++(0.96,0.96);
          \node[text=\tc] at (\c+0.48,-\r+0.48) {\co/\ti};
      }
      \foreach \r/\name in {0/lin-3 remote, 1/lin-4 ladder, 2/lin-4 bidir, 3/lin-4 repeat}
         \node[left] at (-0.05,-\r+0.48) {\name};
      \foreach \c/\o in {0/0,1/1,2/2,3/3}
         \node[above] at (\c+0.48,1.08) {opt \o};
      \node[align=center] at (2,-4.05) {%
        \tikz\fill[black!60](0,0)rectangle(0.28,0.28); contradicted \quad
        \tikz\fill[black!12](0,0)rectangle(0.28,0.28); tie};
    \end{tikzpicture}
    \caption{EX-C7 locked-cell grid. Each cell aggregates five transpiler seeds and reports contradicted/tied counts. No cell preserves the claimed direction.}
    \label{fig:exc7_heatmap}
\end{figure}

\begin{table*}[t]
\centering
\caption{RQ2 audit summary by evidence tier. Rows report audit records, locked designs, tier-specific outputs, and boundary or qualification notes.}
\label{tab:external_audit}
\footnotesize
\setlength{\tabcolsep}{3pt}
\begin{tabular}{@{}p{0.09\textwidth}
>{\raggedright\arraybackslash}p{0.22\textwidth}
>{\raggedright\arraybackslash}p{0.23\textwidth}
>{\raggedright\arraybackslash}p{0.235\textwidth}
>{\raggedright\arraybackslash}p{0.15\textwidth}@{}}
\toprule
\textbf{Audit record} &
\textbf{Claim / comparison object} &
\textbf{Audit design} &
\textbf{Output / status} &
\textbf{Boundary / qualification} \\
\midrule
    \multicolumn{5}{@{}l}{\textbf{Tier 1 --- Proxy-free scoped scalar-directional audit records (Wilson S/U/\rev{NP} verdicts)}} \\
    \addlinespace[2pt]

    EX-C1~\cite{kharkov2022arline} &
    Pytket $<$ Qiskit, compile time &
    4 circuits $\times$ 4 opt levels $\times$ 5 timing reps &
    $80/80 \to$ Sustained &
    L2/L3 metric-irrelevant \\

    EX-C2~\cite{christiansen2024quantum} &
    QAOA approx.-ratio improvement &
    16 instances $\times$ 5 copula depths &
    $80/80 \to$ Sustained &
    aggregate/cluster traceability boundary \\

    EX-C3~\cite{meijer2023comparison} &
    Tket $<$ Qiskit, CNOT count &
    \emph{primary}: 4 circuits $\times$ 4 opt levels $\times$ 5 seeds; \emph{seed exp.}: same $\times$ 20 seeds &
    $62/80 \to$ Unresolved; seed exp. $243/320 \to$ U &
    L2/L3 source-confined \\

    EX-C4~\cite{christiansen2024quantum} &
    QAOA success-prob. improvement &
    5 sizes $\times$ 8 depths $\times$ 2 optimizers &
    $45/80 \to$ Unresolved &
    aggregate/cluster traceability boundary \\

    EX-C5~\cite{nation2024benchmarking} &
    Tket $<$ Qiskit, 2Q depth &
    \emph{primary}: 4 circuits $\times$ 4 opt levels $\times$ 5 seeds; \emph{seed exp.}: same $\times$ 20 seeds &
    $44/80 \to$ Unresolved; seed exp. $180/320 \to$ U &
    L2/L3 source-confined \\

    EX-C6~\cite{kharkov2022arline} &
    Qiskit $>$ Pytket, CX depth &
    \emph{primary}: 4 opt levels $\times$ 20 seeds (fixed graph); \emph{seed exp.}: same $\times$ 80 seeds &
    $34/80 \to$ Unresolved; seed exp. $144/320 \to$ U &
    seed tied to fixed graph scope \\

    EX-C7~\cite{nation2024benchmarking} &
    Qiskit $<$ Tket, 2Q gates &
    \emph{primary}: 4 connectivity-stress circuits $\times$ 4 opt levels $\times$ 5 seeds; \emph{seed exp.}: same $\times$ 20 seeds &
    $0/80$ (62 contra, 18 tie) $\to$ \rev{Not preserved}; seed exp. $0/320 \to$ \rev{NP} &
    source-supported connectivity-stress scope; strict-contradiction-dominated \\

    EX-C8~\cite{nation2024benchmarking} &
    Cirq $<$ Qiskit, build time &
    4 circuit sizes $\times$ 20 timing reps &
    $0/80 \to$ \rev{Not preserved} &
    no ties; strict reversal \\

    \midrule
    \multicolumn{5}{@{}l}{\textbf{Tier 2 --- Source-supported summaries and diagnostic records}} \\
    \addlinespace[2pt]

    SD1~\cite{tan2020optimal} &
    OLSQ $<$ tket, CX count; source rows &
    22 OLSQ source rows; no ENV sweep &
    Source-row summary: $16/22$ preserve, 6 tie &
    tket version per source \\

    SD2~\cite{tan2020optimal} &
    OLSQ $<$ tket, depth \& SWAP; recovered &
    7 recovered comparisons &
    Recovered summary: $7/7$ preserve for depth and SWAP &
    small $N$ \\

    SD3~\cite{osaba2024eclipse} &
    Qrisp $<$ Qiskit-lib, QAOA depth; extension &
    3 sizes $\times$ 5 graphs $\times$ 3 $p$-depths &
    Diagnostic extension: $45/45$ preserve &
    extension, not reported comparison \\

    SD4~\cite{gheorghiu2020reducing} &
    CNOT-OPT $<$ current baseline, CNOT &
    recovered circuits $\times$ 2 architectures (240 cells) &
    Diagnostic surface: $0/240$ preserve &
    baseline-version boundary \\

    SD5~\cite{meijer2023comparison} &
    PauliOpt $<$ TKET, per-device CNOT crossover &
    5 IBM devices $\times$ gadget-size sweep &
    Source-supported crossover summary: $5/5$ devices match the reported pattern &
    derived predicate; limited tail evidence \\

    SD6~\cite{tan2020optimal} &
    OLSQ aggregate CX reduction; geometric-mean rule &
    22 source rows $+$ geometric-mean rule &
    Source-rule summary: geometric mean $0.36$ ($<1$) supports the aggregate relation &
    row-level $16/22$ diagnostic only \\

    SD7~\cite{gheorghiu2020reducing} &
    SD4 baseline-policy sensitivity &
    2 architectures $\times$ 4 Qiskit opt levels &
    Diagnostic sensitivity surface: opt. level 0: $5/50$ \& $5/70$; opt. levels $\geq$1: $0$ &
    sensitivity to SD4 baseline policy \\

    \midrule
    \multicolumn{5}{@{}l}{\textbf{Tier 3 --- Boundary audit records}} \\
    \addlinespace[2pt]

    BD1~\cite{rakyta2022highly} &
    SQUANDER $<$ Qiskit, depth (intended) &
    cannot lock comparison design &
    No verdict &
    OCR-unreliable; $467\times$ vs $479.5\times$ \\

    BD2~\cite{meijer2023comparison} &
    PauliOpt $<$ Qiskit/TKET, runtime (intended) &
    cannot lock timing evidence &
    No verdict &
    uncontrolled host timing; no source runtime \\
    \bottomrule
\end{tabular}

\vspace{2pt}
\footnotesize
S/U/\rev{NP} are Wilson outcomes for Tier-1 proxy-free scoped scalar-directional audit records ($\tau=0.95$). Tier-2 rows report source-supported summaries or diagnostics, not Tier-1 verdicts. No verdict means that the design, evidence object, or outcome rule could not be locked. Seed exp. replays the locked generator with more seeds only; U/\rev{NP} abbreviate Unresolved/\rev{Not preserved}. 2Q denotes two-qubit; CX denotes CNOT. Per-record manifests and \rev{the Tier-1} traces are in the artifact~\cite{claimstab_artifact}.
\end{table*}

\textit{Summary records:}
Tier-2 records report the comparison using the evidence and outcome rule available in the source, such as source rows, recovered comparisons, device curves, aggregate rules, or diagnostic surfaces (Table~\ref{tab:external_audit}). SD6 illustrates why the relation rule matters: the geometric-mean aggregate stated in the source ($0.36 < 1$) supports the reported reduction, while the row-level preservation summary is reported only as diagnostic evidence.

\textit{Boundary audit records:}
BD1 and BD2 cannot lock a comparison design (Table~\ref{tab:external_audit}): one has inconsistent source table values, and the other is a runtime comparison in an uncontrolled timing environment. \framework{} records the evidence boundary and assigns no verdict.

\rqanswer
{\framework{} aligns audit outputs with the available evidence: proxy-free scoped records receive scoped verdicts, source-supported non-Tier-1 records receive source-rule summaries or diagnostic outputs, and unsupported records become explicit evidence boundaries.}

\subsection{RQ3: Can \framework{} explain where comparative support is lost?}
\label{sec:eval_rq3}

RQ3 evaluates whether \framework{} explains where comparative support is lost once a comparison is audited over locked, source-aligned axes. We use three controlled mechanism checks: (1) a C1--C24 canonical diagnostic contrast over benchmark-relevant quantum workloads and problem settings; (2) axis-controlled diagnostic surfaces that isolate L1, L2, L3, ALGO, and combined stack effects; and (3) decision-rule controls for the Wilson S/U/\rev{NP} policy under known preservation rates and reversed cases.

\textit{Canonical diagnostic contrast:}
The C1--C24 contrast contains 24 controlled comparisons from eight benchmark-relevant base setups. C1--C9 are structural compiler-output comparisons over Arithmetic, GHZ, and QFT circuits, comparing pytket and Qiskit on two-qubit gate count. Such circuit families and structural metrics are common in quantum compiler and circuit-benchmarking practice~\cite{nation2024benchmarking,kharkov2022arline,li2023qasmbench,quetschlich2023mqtbench}. C10--C24 are QAOA layer-count comparisons over Max-2-SAT and MaxCut instances, comparing $p{=}2$ and $p{=}1$ at fixed angles $\gamma=0.8$ and $\beta=0.4$; MaxCut and 2-SAT are standard QAOA benchmarking problems, and MaxCut depth comparisons beyond $p{=}1$ have been studied directly~\cite{farhi2014quantum,willsch2020benchmarking,wurtz2021maxcut}. The contrast compares three simpler audit protocols with the full locked audit design. A single-configuration run reports only a direction and matches the full-audit direction in $19/24$ records, while opposing it in $5/24$. A seed-only audit calls all 24 records Unresolved and agrees with the full audit on $14/24$. A single-axis L1 audit is available for 15 records and disagrees with the full audit on $10/15$. The full locked audit design yields $10$ Sustained, $14$ Unresolved, and $0$ \rev{Not preserved} within audited scope. Thus, on canonical quantum workloads, simpler protocols can preserve an apparent direction while missing the audit-axis combinations that weaken comparative support.

\textit{Axis-controlled diagnostic surfaces:}
The controlled diagnostic suite operationalizes audit-axis groups through finite, predeclared grids while keeping the comparator pair, metric, relation rule, and inactive axes fixed. It covers five surfaces: L1 varies within-Qiskit compilation settings over structural circuits; L2 varies shot count and simulator seed for fixed-angle QAOA on deterministic MaxCut graphs; L3 varies six simulator-noise regimes over the same QAOA comparison; ALGO varies optimizer family and initialization seed under a fixed iteration budget; and a combined surface jointly varies L1, L2, and L3 on a smaller MaxCut pool. Across the 15 controlled-surface records, the suite yields $0$ Sustained, $13$ Unresolved, and $2$ \rev{Not preserved} within audited scope. The two \rev{Not preserved} outcomes are tie-driven L1 structural cases \rev{(all 1,200 cells tied in each)}, while L2, L3, ALGO, and combined surfaces remain Unresolved under the locked grids.

\textit{Example controlled diagnostic:}
One controlled MaxCut-QAOA diagnostic audits the direction $p{=}2 > p{=}1$, a QAOA layer-count comparison studied in MaxCut analyses~\cite{farhi2014quantum,wurtz2021maxcut}. Under a combined L1$\times$L2$\times$L3 stress design with $\delta=0.05$, the direction is preserved in 723 of 1,440 locked cells ($\hat{s}=0.502$, Wilson 95\% interval [0.476, 0.528]), yielding an Unresolved outcome. The apparent layer-count advantage therefore becomes unsupported once compilation, sampling, and noise axes are inspected together.

\textit{Decision-rule controls:}
We evaluate the scalar-directional decision rule with two controls. First, a Wilson calibration grid fixes a true preservation rate $s_{\mathrm{true}}$ and locked-design size $N$, samples binary direction-preservation indicators, and applies the S/U/\rev{NP} rule to the resulting preservation count. The grid contains 18 profiles, crossing $s_{\mathrm{true}}\in\{0.50,0.70,0.85,0.95,0.99,1.00\}$ with $N\in\{10,30,100\}$; across 3000 trials, the Wilson 95\% interval contains the true preservation rate in 94.97\% of trials. Second, a reversed-detection control \rev{labels} 26 of 27 known-reversed cases \rev{Not preserved}. The artifact provides the diagnostic grids, per-case audit-space construction, per-diagnostic summaries, and verification pointers for RQ3~\cite{claimstab_artifact}.

\rqanswer
{On benchmark-relevant quantum workloads and controlled audit surfaces, \framework{} exposes how apparent comparative directions can weaken or become Unresolved when source-aligned stack axes are locked and inspected.}

\subsection{RQ4: Are verdicts robust to statistical and declared audit choices?}
\label{sec:eval_rq4}

RQ4 checks whether scalar-directional verdicts depend on statistical choices or declared audit choices. We treat S/U/\rev{NP} labels as scoped classifications over locked audit designs, then evaluate sensitivity to preservation threshold $\tau$, dependence-aware grouping, tie handling, budget or seed-count expansion, practical-effect threshold $\epsilon$, and ALGO-axis bookkeeping when the required traces are available. Table~\ref{tab:robustness_summary} summarizes the checks.

\begin{table}[t]
\centering
\caption{Verdict-level robustness checks.}
\label{tab:robustness_summary}
\scriptsize
\setlength{\tabcolsep}{2.5pt}
\renewcommand{\arraystretch}{1.08}
\begin{tabularx}{\columnwidth}{@{}p{0.24\columnwidth}p{0.22\columnwidth}X@{}}
\toprule
\textbf{Robustness check} & \textbf{Scope} & \textbf{Result and interpretation} \\
\midrule
Threshold $\tau$ &
Tier-1 scoped + 24 canonical &
Tier-1 labels are unchanged at $\tau\in\{0.90,0.95\}$; at $\tau=0.99$, finite-$N$ extreme labels soften to Unresolved. The canonical check shifts 10 borderline cases to Unresolved. \\
\addlinespace
Dependence-aware grouping &
\rev{RQ3 claims} + Tier-1 per-cell records &
The \rev{RQ3} grouping check \rev{($B=1{,}000$)} agrees with the primary classifications. Tier-1 per-cell checks \rev{($B=10{,}000$)} identify EX-C3 as cluster-bootstrap-sensitive; EX-C2 and EX-C4 are unavailable for this check. \\
\addlinespace
Tie handling &
Strict scalar-directional per-cell records &
EX-C7 has no preserved cells and remains contradicted-dominated, but its categorical \rev{Not preserved} label is sensitive to alternative tie handling; records without per-cell traces are unavailable. \\
\addlinespace
Budget / seed-count expansion &
Replayable Tier-1 compiler records &
All replayable compiler records keep their primary labels through $N=320$; EX-C7 remains at 0/320 preserved cells. \\
\addlinespace
Practical-effect $\epsilon$ and ALGO bookkeeping &
Applicable diagnostic records &
No classification changes among analyzable practical-effect checks; ALGO bookkeeping does not change matched records or Wilson labels. \\
\bottomrule
\end{tabularx}
\end{table}

The checks preserve the primary verdicts under the predeclared audit settings and identify the remaining boundary cases. EX-C3 is cluster-bootstrap-sensitive; single-cluster deletion keeps its primary Unresolved label unchanged. EX-C2 and EX-C4 retain their primary verdicts; per-cell dependence and tie-handling checks are unavailable because the required traces are not materialized. EX-C7 is contradicted-dominated: no locked or expanded cell preserves the reported direction, including $0/80$ primary cells and $0/320$ cells under locked seed-count expansion. Under stricter threshold or alternative tie-handling policies, the categorical \rev{Not preserved} label is marked as policy-sensitive, even though the observed preservation count remains zero.

\rqanswer
{Stable outcomes are grounded in locked evidence and declared audit choices, while sensitivity and missing traces are reported as explicit evidence boundaries.}


\textit{Evidence synthesis:}
Together, the four RQs show that reported comparisons in empirical QSE require auditing the comparison itself alongside execution-level reproducibility. \framework{} makes this audit inspectable by identifying what can be materialized, what the locked evidence supports, and where the evidence boundary lies.

\section{Threats to Validity}
\label{sec}
This work is subject to \rev{five} main threats to validity, concerning decision-rule dependence, human judgment, corpus and diagnostic scope, hardware/toolchain scope\rev{, and domain scope}.

\textit{Decision rule and dependence:}
S/U/\rev{NP} labels are scoped decisions over locked comparison records, not claims about all possible executions. Audit cells may share circuits, instances, seeds, optimization levels, timing groups, or environment settings. We therefore interpret verdicts within the locked finite design and report sensitivity checks when traces permit; unsupported checks are recorded as evidence boundaries.

\textit{Human judgment:}
Claim extraction, claim-card construction, admissibility decisions, and evidence-boundary annotations require judgment. We mitigate this threat through the coding protocol and inter-auditor checks in Section~\ref{sec:eval_rq1}, including blinded recoding, stratified consistency auditing, and adjudication against the source paper and artifact. The artifact records decisions, exclusions, axis labels, missing-evidence reasons, and adjudication records~\cite{claimstab_artifact}.

\textit{Corpus and diagnostic scope:}
The corpus is Google Scholar-based and reflects a sampled corpus. Relevant comparisons outside the indexed or screened results, or under different terminology, may be missed. The materialization funnel is therefore evidence for the sampled corpus. The eight proxy-free scoped records support scoped S/U/\rev{NP} outcomes over materialized records, while their small number is part of the materialization-gap result. Controlled diagnostics use benchmark-relevant quantum-workload cases and axis-controlled surfaces, including structural circuit-compilation and MaxCut/Max-2-SAT QAOA comparisons, to expose mechanisms of comparative-support loss under fixed audit axes; they are not prevalence estimates over all empirical QSE claims.

\textit{Hardware and toolchain scope:}
The L3 evidence in this study is limited to controlled simulator-noise regimes on Qiskit Aer and Qiskit fake backends. Live-hardware drift, cross-provider variation, queue effects, calibration updates, and future toolchain changes are outside the locked audit records. Compiler and runtime verdicts are scoped to the locked toolchain version and recorded timing environment.

\rev{\textit{Domain scope:} The audit logic (claim cards, locked designs, the decision rule, and evidence boundaries) could be applied to empirical comparisons in software engineering generally. The need for such audits is greater in quantum software: execution is noisy, and comparative results shift with compilation, sampling, backend noise, and stochastic optimization~\cite{preskill2018quantum,dasgupta2024stability}. This study therefore focuses on QSE, and its axis groups and metrics are specific to this domain.}

\section{Related Work}
\label{sec:related}

This work is closest to robustness analysis, quantum-software testing, benchmarking, reproducibility infrastructure, and platform-stability studies. These areas provide adjacent evidence; \framework{} audits the reported comparison itself.

\textit{Specification-space and robustness analyses:}
Multiverse analysis~\cite{steegen2016increasing} and specification-curve analysis~\cite{simonsohn2020specification} test whether an empirical finding survives alternative defensible analysis choices. \framework{} shares this robustness intuition, but its admissible space is source-bounded, not analyst-controlled: baselines, circuits, seeds, budgets, backend settings, and outcome rules can vary only when they are source-supported, metric-affecting, and relation-preserving.



\textit{Quantum-software testing and benchmarking:}
MorphQ~\cite{paltenghi2023morphq}, QDiff~\cite{wang2021qdiff}, Muskit~\cite{mendiluze2021muskit}, Quito~\cite{wang2021quito}, MorphQ++~\cite{kitt2024morphq++}, and QITE~\cite{paltenghi2025qite} test programs, platforms, or implementations. Benchmarking suites such as Benchpress~\cite{nation2024benchmarking}, Arline~\cite{kharkov2022arline}, QASMBench~\cite{li2023qasmbench}, MQT Bench~\cite{quetschlich2023mqtbench}, and QSimBench~\cite{bisicchia2025qsimbench} produce measurements from which comparative claims are often drawn. \framework{} instead checks whether the reported comparison is materializable and preserved under the source-supported comparison design.



\textit{Reproducibility, artifacts, and platform variation:}
Empirical software engineering and quantum-software reproducibility work asks whether artifacts can be accessed, executed, reused, or rerun~\cite{muttakin2026state,mauerer20221,gierisch2025qef,moguel2025quantum,dasgupta2021reproducibility}. Hardware-stability and platform studies explain why quantum-software evidence varies across time, platforms, and toolchains~\cite{dasgupta2024stability,baheri2022quantum,dasgupta2023adaptive,paltenghi2022bugs,de2024quantum,garcia2023quantum}. \framework{} uses this insight to separate admissible variation from unsupported variation, which becomes an evidence boundary.
\section{Conclusions}
\label{sec:conclusion}

Empirical QSE comparisons are often reported as performance orderings, but our audit \rev{of 455 claims from 119 sampled papers} shows that the ordering and its supporting evidence are separable. A comparison can be identifiable, representable, and even partially reproducible while still lacking the matched baselines, settings, instances, budgets, conventions, or execution records needed for proxy-free relation-level auditing. \framework{} makes this distinction explicit by reporting either a scoped relation outcome or an evidence boundary.

The main lesson is that \rev{in our sample,} many empirical quantum-software comparisons are reported as outcomes without the evidence needed to audit the comparison itself. For empirical QSE, this shifts the reporting target: sources should state the relation being claimed, expose the evidence needed to lock that relation, and clearly mark where the available evidence no longer supports the comparison.


\section{Data Availability}
\rev{The artifact is archived at \url{https://doi.org/10.5281/zenodo.23659820}~\cite{claimstab_artifact}; the development repository is \url{https://github.com/Bossy-Ye/artifact-review-claimstab}.}

\section*{Acknowledgments}
\rev{We thank the anonymous reviewers for their thoughtful and careful reviews. This work has been partially supported by NSFC with Grant Nos. 92582203 and 62672372.}

\bibliographystyle{IEEEtran}
\bibliography{references}

\end{document}